\documentstyle[aaspp4,epsf,11pt]{article}
\tighten
\begin{document}
\lefthead{Pavlov et al.}
\righthead{Surfaces of Isolated Neutron Stars}
\title{Multiwavelength observations of isolated neutron stars
as a tool to probe the properties of their surfaces}
\singlespace

\author{G.~G. Pavlov}
\affil{The Pennsylvania State University, 525 Davey Lab,
University Park, PA 16802, USA; pavlov@astro.psu.edu} 
\and 
\author{V.~E. Zavlin,
J. Tr\"umper, and
R. Neuh\"auser}
\affil{Max-Planck-Institut f\"ur Extraterrestrische Physik, D-85740 
Garching, Germany; zavlin@rosat.mpe-garching.mpg.de}
\begin{abstract}
We show that an analysis of multiwavelength observations
of isolated neutron stars based on neutron star atmosphere
models can be used not only to evaluate the neutron star effective
temperature, but also to determine chemical composition of its 
surface.  To demonstrate how this new method can be applied to
a specific object, we chose the old isolated neutron star 
candidate RXJ1856.5--3754, whose soft X-ray radiation
has been studied recently by Walter, Wolk, \&
Neuh\"auser (1996). We fitted the soft X-ray spectrum
of this object with neutron star atmosphere models
of different chemical compositions and used these fits
to calculate the source spectrum over a broad wavelength
range. We showed, in particular, that
the optical/UV flux expected from this object 
depends drastically on the composition of its surface. In particular,
the neutron star covered with hydrogen would be 5--6 magnitudes
brighter than the neutron star with an iron surface.
The object should also be observable with $EUVE$;
the EUV flux is expected to be almost twice higher for the iron
surface than for the hydrogen one. 
Thus, multiwavelength observations of this object would enable one
to examine, for the first time, chemical composition of the
neutron star surface.
The method proposed can be applied to other nearby
isolated neutron stars.
\end{abstract}
\section{Introduction}
It follows from both extrapolating 
pulsar birthrates (Narayan \& Ostriker 1990)
and from the number of supernovae required to account
for the heavy element abundance (Arnett et al. 1989)
that the Galaxy is populated by about $10^{8}-10^{9}$ neutron stars (NSs). 
However, only $\sim 10^{-6}-10^{-5}$ of this amount 
has been observed so far, mainly as radio pulsars.
Thus, one can expect that there exists
a vast population of nonpulsating NSs which
could be observed as faint sources of thermal radiation.
It has recently been claimed by Walter, Wolk, \& Neuh\"auser 
(1996; hereafter WWN96) that the 
bright, nonpulsating, soft $ROSAT$ 
X-ray source RXJ1856.5--3754 is a viable candidate to be a member
of this population.

WWN96 found that (a) a blackbody fit of 
the spectrum detected with the
$ROSAT$ Positional Sensitive Proportional Counter (PSPC) 
(count rate $3.62\pm 0.03$ s$^{-1}$) yields an apparent blackbody
temperature of $6.6\times 10^5$ K, a hydrogen column density
of $1.4\times 10^{20}$ cm$^{-2}$,
and an emitting area of $480\, (d/100\,{\rm pc})$ km$^2$;
(b) other standard models 
do not give better fits 
(see also Neuh\"auser et al. 1996 for details);
(c) the $ROSAT$ High Resolution Imager (HRI)
count rate is $0.56\pm 0.005$ s$^{-1}$;
(d) there is no evidence for X-ray variability in this source;
and (e) the object is fainter than $V\simeq 23$ in the visual band.
No radio, UV, or gamma radiation have been detected.
The source is projected onto the R CrA dark molecular cloud, at a
distance of $\sim 120$ pc (Wang 1994). The observed X-ray extinction
allows one to assume that the source is closer than the cloud.

The inferred temperature looks surprisingly
high for this presumably very old NS (see arguments in
WWN96), although it may be caused by enhanced
accretion of interstellar matter if the NS
relative motion is slow enough.  However, the
real spectrum of the NS thermal emission may differ
substantially from the Planck spectrum due to the presence
of an atmosphere on the NS surface (e.~g., Pavlov et al.~1995),
and the actual shape of the spectrum strongly depends on
chemical composition of the emitting layers 
(Rajagopal \& Romani 1996; 
Zavlin, Pavlov, \& Shibanov 1996, hereafter ZPS96).
Fitting  observed spectra with the model atmosphere spectra 
may result in 
quite different effective temperature
$T_{\rm eff}$ and
ratio $R/d$ of the size of emitting region to the distance.

The chemical composition of the NS surface depends on
various effects.
Due to strong gravitation at the NS surface, the elements 
there are separated in accordance with their
atomic weights so that the lightest
element resides at the very surface and forms the atmosphere, 
whereas heavier elements sink down.  
Thus, if an amount of hydrogen (H) is present in the outer
layers (e.~g., due to accretion of the interstellar
matter), one can expect a pure H atmosphere.
If during the NS life time there were no substantial
accretion of the H-rich matter, or the hydrogen
were burnt or ejected from the surface, the atmosphere
may be comprised of a heavier element. In principle,
it might be helium (He) or carbon, although the most natural choice
seems to be iron (Fe) formed in the core of the supernova
progenitor star. The H and He atmospheres
emit spectra which are much harder in the Wien tail
 than the blackbody (BB) spectrum because hotter deep layers are seen
at high frequencies where the spectral opacity is lower.
In this case, the best-fit atmosphere effective temperature
is considerably lower than the BB temperature, whereas the 
$R/d$ ratio  is larger than for the BB
fit. On the contrary, in the case
of the Fe atmosphere the overall shape of the spectrum
is not so drastically different from that of the  BB model,
although local deviations may be quite substantial due to numerous
spectral lines and photoionization edges. 

The chemical composition of the NS surface could 
be determined by the analysis of its soft X-ray spectrum.
However, spectral resolution and sensitivity of 
current soft X-ray detectors is not high enough to firmly detect
spectral features in the most interesting
range $\sim 0.1-1$ keV.  On the other hand, 
this range is so short that the shape of the
continuum can be fitted with quite different
models. Thus, 
it is desirable to measure the NS flux over a broader range.
Since the NS thermal flux
is intrinsically very low at medium and hard X-rays,
and it is severely suppressed by the interstellar extinction
at EUV, the most perspective would be optical/UV observations
($E\simeq 2-10$ eV). Although NSs are extremely faint in this
range (typical expected magnitudes 
are $\sim 26-28$ at $d\sim 100$ pc),
broad-band photometry
of these sources is quite feasible with the {\em Hubble
Space Telescope} ({\em HST}) and the largest
ground-based telescopes (Pavlov 1992).

So far, optical/UV {\em thermal} radiation has been
apparently detected from only two active pulsars, PSR B1929+10 and
(more questionable) PSR B0950+08 (Pavlov et al. 1996); 
optical radiation from younger,
`middle-aged' pulsars Geminga and PSR B0656+14 
is most likely of 
a non-thermal origin (Bignami et al. 1996; Pavlov et al. 1996).
The pulsars B1929+10 and B0950+08 have relatively strong magnetic
fields ($\sim 10^{12}$ G), which precludes analysis of the
atmosphere chemical composition because no reliable opacities
have been calculated for elements other than hydrogen for
strongly magnetized dense plasmas. On the contrary, for very old
NSs, where the magnetic field is believed to have been decayed
and rotation has slowed down
so that the NSs cannot manifest themselves as active pulsars, 
the chemical composition can be analyzed with the use of
non-magnetic atmosphere models. 

In the present paper we 
fit the $ROSAT$ PSPC spectra 
of the NS candidate RXJ1856.5--3754, which seems to be sufficiently old
and does not show pulsar activity,
with the H, He, and Fe
 atmosphere models developed by ZPS96.
We use these fits to calculate the IR through X-ray fluxes expected
from this source, 
assuming that it is indeed an old, isolated
NS with low magnetic field. We show that the detected
optical flux (and the distance to the object)
would immediately allow us to distinguish
between light-element (H or He) and heavy-element composition.
We show that for any assumptions
on the chemical composition the object should be observable with
the $EUVE$ Deep Survey (DS) instrument in a relatively short
exposure time.  The most dramatic effect of the chemical
composition is expected at  
optical wavelengths, where
the NS covered with hydrogen would be about $5-6$ magnitudes brighter
than the NS with an iron surface.
\section{PSPC fits and the sourse flux in a broad energy range}
We re-analized the PSPC data using 
the EXSAS software (Zimmermann et al. 1994).
The source in the PSPC image is $14\farcm 6$ off-axis, and its
 measured (off-axis) count rate is  $3.23
\pm 0.04$ ${\rm s}^{-1}$,
which corresponds to the corrected (on-axis) 
count rate of $3.64\pm 0.04$ s$^{-1}$,
in agreement with WWN96.

We binned the observed count rate spectrum into 19 energy channels 
between approximately $0.1$ and
$1.0$ keV (no source counts were detected above 1 keV)
and fitted it with the H, He, and Fe atmosphere models
of ZPS96, and with the BB spectrum. The best-fit parameters
(effective hydrogen column density $n_H$, 
unredshifted surface effective temperature
$T_{\rm eff}$, distance to the source $d$),
and reduced $\chi_\nu^2$ (for 16 degrees of freedom)
are presented in Table 1. We assumed 
standard NS mass $M=1.4 M_\odot$ and radius $R=10$ km 
(which corresponds to
 the apparent radius $R_\infty =g_r^{-1} R  =13$~km,
where $g_r= [1-(2GM/Rc^2)]^{1/2}$).
Although the BB temperature and $n_H$ are essentially the same
as obtained by WWN96 (notice that the unredshifted
temperature $T_{\rm eff}$ is higher  than the apparent
temperature $T_{\rm eff,\infty}=g_r T_{\rm eff}$
presented by WWN96), the formal quality of our fit is much better
(minimum $\chi_\nu^2 = 0.80$ vs. $2.31$), most likely
 because we chose another binning of the PSPC count spectrum.

\begin{table}[t]
\caption{PSPC fits, corresponding $V$ magnitudes, and count rates
for other detectors}
\begin{tabular}{ccccccccc}
\tableline
\tableline
Model & $n_H$ & $T_{\rm eff}$ & $d$ & $\chi_\nu^2$ & V & 
HRI\tablenotemark{a} & DS\tablenotemark{b} & Scanner\tablenotemark{c} \\
 & ($10^{20}$ cm$^{-2}$) & ($10^5$ K)  & (pc)  &  & mag & (s$^{-1}$)  &(s$^{-1}$) &(s$^{-1}$) \\ 
\tableline
H   & $2.14 \pm 0.08$      & $2.04 \pm 0.36$  & $4.8 \pm 1.1$ & 0.94 & 21.8 &
 0.60 &
0.046 & 0.045 \\
He  & $2.30 \pm 0.08$    & $1.95 \pm 0.31$   & $3.8 \pm 1.0$ & 1.24 & 20.8 &
 0.60 & 0.043 & 0.043 \\
Fe & $0.95 \pm 0.04$  & $9.78 \pm 1.45$  & $216 \pm 33$ & 0.68 & 27.5 & 0.69 & 
0.087 & 0.065 \\
blackbody & $1.49 \pm 0.03$ & $8.51 \pm 1.43$ &  $158 \pm 26$ & 0.80 &
26.8 & 0.61 & 0.053 & 0.047 \\
H magnetic\tablenotemark{d} & $2.21\pm 0.07$ & $2.88\pm 0.43$ & $12.1\pm 1.3$ & 1.32 &
23.3 & 0.60 &
0.044 & 0.043  \\
\tableline
\tablenotetext{a}{Predicted $ROSAT$ HRI count rate}
\tablenotetext{b}{Count rate predicted for the $EUVE$ Deep Survey instrument
with the Lexan/B filter}
\tablenotetext{c}{Count rate predicted for the $EUVE$ Scanner A/B with
the Lexan/B filter}
\tablenotetext{d}{For $B=1\times 10^{12}$~G}
\end{tabular}
\end{table}

%

Since the spectra emitted by the H and He atmospheres 
are substantially harder, in the $ROSAT$ range, than the BB
spectrum with the same effective
temperature, we obtained quite different fitting parameters
for the atmosphere model fits.
For example, $T^{\rm BB}_{\rm eff}/T^{\rm H}_{\rm eff}=4.17$ and 
$d^{\rm BB}/d^{\rm H}=32.8$.
On the other hand, the difference between the parameters
obtained for the BB and Fe atmosphere fits is not so dramatic because
the Fe atmosphere spectrum is close, on  average, to the BB
spectrum of the same temperature (ZPS96). Moreover, 
the Fe model spectrum is even slightly softer than BB
in the $ROSAT$  range  due to strong 
bound-free absorption, which leads to higher $T_{\rm eff}$ and $d$.
Despite the significant difference in the best-fit parameters 
(particularly, $T_{\rm eff}$ and $d$), the fits look very similar. 
From the viewpoint of formal
statistics, all the four fits are acceptable, i.~e., one cannot
choose the most acceptable model without invoking additional
arguments.

The projections of the three-dimensional confidence regions
onto the $T_{\rm eff}$-$n_H$ plane, together with the lines
of constant most probable distances, 
are shown in Figure 1 for the four spectral models.
We see that the confidence regions for 
H and He models overlap with each other,
being very distant from those for the Fe and 
BB models.

\epsfxsize=17cm
\epsffile[10 130 800 550]{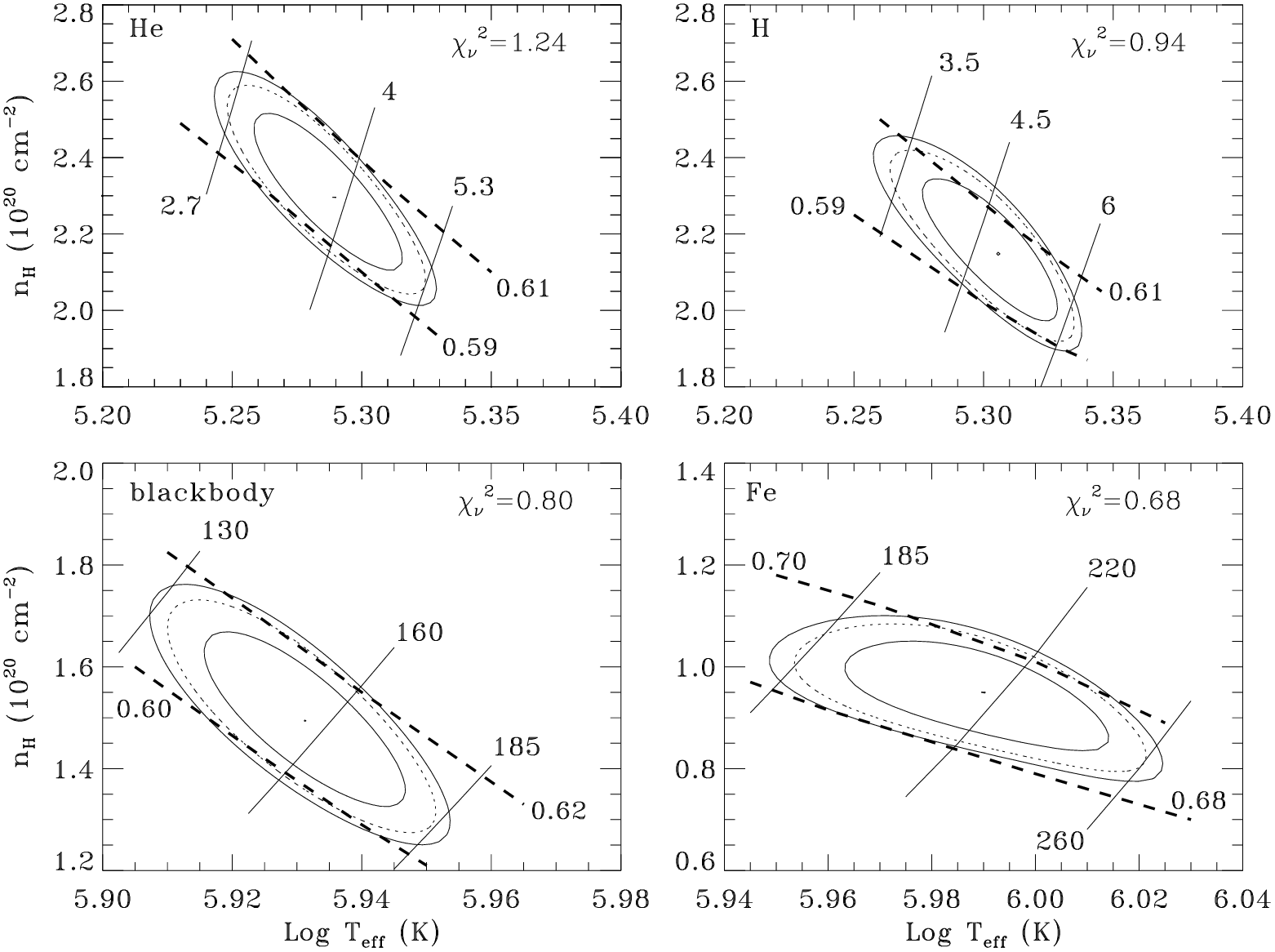}

\begin{small}
{\bf Fig.~1.}
Confidence contours for the helium, hydrogen, and iron atmosphere
models, and the blackbody model, at 68\% ($1 \sigma$), 90\% and 99\%
 confidence levels.
Nearly straight solid lines are the lines of constant most probable
distance $d$ (in pc). The thick dashed lines correspond to
the constant HRI count rates (figures near the curves, in s$^{-1}$).
\end{small}
\bigskip

Since the energy dependence of the HRI 
and PSPC effective areas are different,
the HRI observations of the same object can provide an additional
opportunity to examine the models for its radiation. 
We re-analyzed the HRI image and obtained the source
count rate $0.56\pm 0.01$ s$^{-1}$.
The HRI count rates corresponding to the best PSPC fit models
are presented in Table 1, and the curves of constant 
 HRI count rates
are shown in Figure 1. We see that the model HRI count rates
exceed the measured one by $5-9\%$ for the H and He models,
$7-11\%$ for the BB model, and $20-25\%$ for the Fe model.
The systematic uncertainty of the HRI effective area at low energies,
$0.1-0.5$ keV, which give the main contribution to the HRI count rate,
may reach $5-7\%$ (B.~Aschenbach, private communication).
Since the PSPC response is also not free from systematic errors
at low energies, one cannot exclude that the 
H and He atmosphere models are consistent with the HRI data,
whereas the Fe  model seems to give a too high HRI count rate.

To demonstrate how observations of this object in other spectral bands
can be used to discriminate between
the source models,  we calculated the spectral fluxes,
both attenuated and non-attenuated by the interstellar extinction,
for the best PSPC fit models 
over the broad photon energy range, $1-1000$ eV (Fig.~2).
To calculate the extinction at IR/optical/UV energies,
we use the approximate relation between the hydrogen
column density and the color excess: $E(B-V)=n_H/(5.8\times 10^{21}
{\rm cm}^{-2})$ --- see Savage \& Mathis (1979).  In the range
of $0.1-1.0$ keV all the attenuated spectra are very similar 
to each  other, especially for the H, He, and BB
models, in accordance with approximately the same quality of the PSPC
fits.  In the range of $\simeq 10-80$ eV the fluxes are unobservable
for all the models because of the interstellar absorption.
They are observable in the high-energy part of the $EUVE$ spectral
range, $\simeq 100-200$ eV, which considerably overlaps with
the $ROSAT$ soft channels.  We evaluated expected count rates
for the $EUVE$ DS 
and $EUVE$ Scanners A/B with the Lexan/B filter for different
models (see Table 1). The count rates are not much different
for the H, He, and BB models, $\simeq 0.04-0.05$
s$^{-1}$, whereas for the Fe model the 
count rate is higher, particularly for the DS instrument.
A $3\sigma$ upper limit of 0.066 s$^{-1}$,
reported by WWN96 for a short (1,001 s) Scanner exposure,
is too high to discriminate between the models
(although the Fe atmosphere count rate almost coincides
with the limit). A longer, $\sim 20 - 30$ ksec, observation
with the $EUVE$ DS is needed to detect $\sim 1000$ source
counts and to draw a convincing conclusion on the source nature.

\epsfxsize=17cm
\epsffile[10 30 800 520]{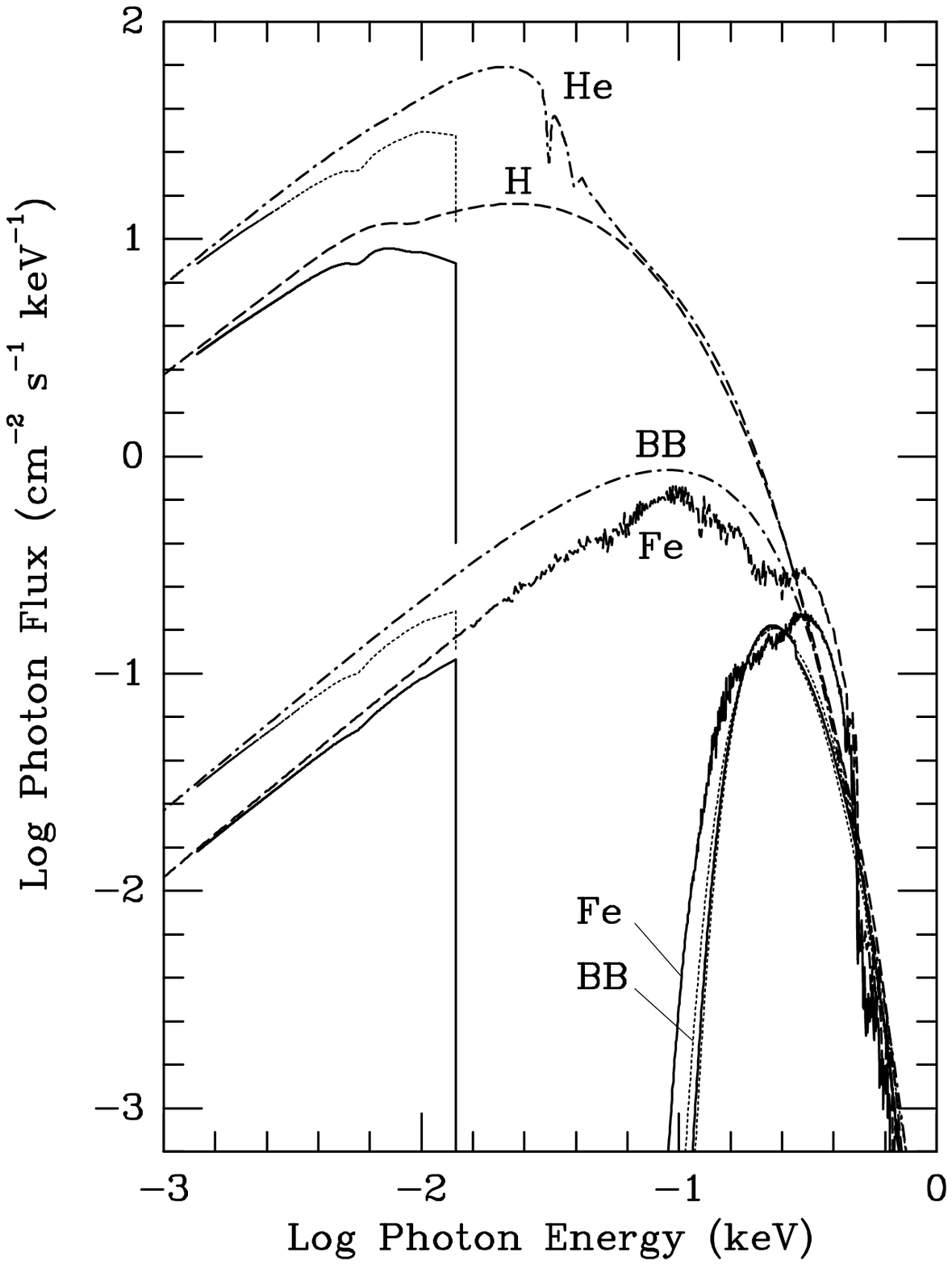}

\begin{small}
{\bf Fig.~2.}
Photon spectral fluxes for four models in the range of $1-1000$ eV with
and without allowance for interstellar absorption.
\end{small}
\bigskip

Much more striking difference between the model fluxes
is seen in the IR/optical/UV range (1--10 eV): for instance,
the source flux is $6-7$ magnitudes brighter for
the H and He models than for the Fe model (see Table 1).
The reason is that a $40-50$ times smaller distance (at a given emitting
area) is needed for the H/He models to provide the same soft-X-ray 
flux at $4-5$ times lower effective temperature. The spectra
at the low energies 
follow the Rayleigh-Jeans law, $F_\nu \propto \nu T (R/d)^2$, 
so that the flux for the H/He models is $400-500$ times
higher than for the Fe model.
Since the effect of extinction is not very significant, 
the source is expected to be the brightest at UV frequencies.
\section{Discussion and conclusions}

The $V$ magnitude predicted for the H/He
models is about 20--22, which is
brighter than the limit of $V=23$ reported by WWN96. If this limit is
correct, then the non-magnetic H/He models can be excluded.
One more striking feature of the H/He atmosphere models
is a very small $d/R$ ratio ---
for instance, for $R=10$ km the distance
would be in a range of $3.1 -6.3$  pc, at a 99\% confidence
level, if the NS is covered with hydrogen.
For the expected local NS density, $\sim 10^{-4} - 10^{-3}$
pc$^{-3}$, the probability to find a NS at such a distance is
low (e.~g, the probability to find one NS at a distance of
5 pc from the Sun is $\sim 0.05-0.5$).
In principle, the NS radius may be larger than 10 km; for example,
$R\simeq 15$ km for a NS of the mass $1.3 M_\odot$ for
a hard equation of state (e.~g., Shapiro \& Teukolsky 1983).
In this case, the distance would become larger by a factor of 1.5.
The distances following from the H/He
atmosphere models can be easily measured
with the $HST$ even for such a faint object.
On the contrary, if we adopt the Fe atmosphere model,
the distance of $180-270$ pc would be even larger than for the
BB fit, considerably exceeding
120 pc to the R CrA molecular
cloud. This also seems unlikely because the total gas column
density of the cloud in this region, $\sim 5\times 10^{20}$
cm$^{-2}$, is substantially greater than
 the 99\% upper limit of $n_H \simeq 1.1 \times 10^{20}$ cm$^{-2}$.

Thus, at the present stage, the situation looks  puzzling
--- light element models give uncomfortably small distances and
optical magnitudes brighter than the reported limit,
whereas the Fe model results in a distance larger than acceptable,
if the NS is  foreground to the R CrA dark cloud.
This conclusion, however, is based on the assumption that the
magnetic field at the 
NS surface does not exceed $10^9-19^{10}$~G (otherwise the
non-magnetic atmosphere models are inapplicable).
Although there have been many speculations about the field decay
in isolated NSs (e.~g., Itoh et al.~1995, and references therein),
this idea is not firmly supported by observations to date. The main
argument against the presence of high magnetic fields in RXJ1856.5--3754
is that no significant pulsations have been found
in the $ROSAT$ data (Becker 1996).
If the lack of pulsations is confirmed, it still might be consistent
with the presence of high magnetic fields if the rotation
axis coincides with the magnetic axis or with
the line-of-sight.  The former is not unexpected for an old NS;
the probability of the latter is very low.
Fitting the observed spectra with the magnetic H atmosphere models
may not be as definitive as with the nonmagnetic ones because
we know nothing about the strength and geometry of the
field which, in turn, affects the distribution of
temperature over the NS surface. Besides, the available magnetic models
may  not be accurate at relatively low temperatures (Pavlov et al.~1995).
An example of the magnetic H atmosphere fit
(assuming uniform temperature and $B=1\times 10^{12}$~G)
is presented in Table 1. Generally, magnetic fits result in
higher effective temperatures, greater distances, $d\sim 10-30$~pc, and
fainter optical magnitudes, $V\simeq 23-25$, than the nonmagnetic
H/He atmosphere fits. Although the distances still look very small,
such parameters do not contradict to the data available. 

The most reliable way to understand the nature of RXJ1856.5--3754
is to conduct deep optical/UV
observations of a relatively wide field around its X-ray position.
It should be noted that although the X-ray position of 
a star detected 
in the HRI pointed observation is coincident, within $0\farcs 7\pm 
2\farcs 6$, with its $HST$ GSC position
(WWN96), the latter may itself be in error by a few arcseconds (e.~g.,
because of a large proper motion).  Hence, the HRI position 
of the NS may be less accurate than $2''$, as it was adopted by WWN96. 
In addition, if the distance to the NS is indeed as low as a few
parsec, one should expect a large proper motion of the object.
For instance, if one adopts the NS tangential velocity of
300 km/s, a typical value for radio pulsars, and a distance
of 10 pc, then the proper motion  would be
$\mu=6\farcs 2\, v_{300}\, d_{10}^{-1}$ yr$^{-1}$.
The proper motion might be limited by 
the relatively small mean offset, $\simeq 26''$, between the
$Einstein$ IPC slew survey position and the HRI position
measured 14 years later. This limit, however, is not very stringent
because of the large uncertainty, $72''$ at 90\% confidence level, 
of the $Einsten$ position:
the resulting proper motion can be estimated as $\mu = 2''
\pm 5''$ yr$^{-1}$.  Thus, one cannot exclude that the object could travel 
$\sim 10''$ during 1.4 years between the HRI and optical observations,
so that the reported limit of $V=23$
should be considered with caution, and a wider field
should be carefully investigated before making final
conclusions. To prove that a faint object
in the field is indeed the NS, the color of the object must be examined
--- NSs of expected temperatures should be very blue.
 
If a more accurate investigation of the field results in detection of
this source with a $V$ magnitude of about $21-22$,
and a distance of $3-7$ pc,
the conclusion would be quite unambiguous: the NS has
an H or He atmosphere and a very low magnetic field. If the
magnitude is slightly fainter, about 24, and the distance larger, $10-30$ pc,
we would argue for the magnetic H atmosphere model and conclude
that magnetic and rotation axes coincide with each other, or
the rotation axis is co-aligned with the line of sight.
Finally, if the magnitude is as faint as $V\simeq 27-28$ and/or
the distance exceeds $\sim 150-200$ pc,
then we conclude that the atmosphere consists of Fe or another 
heavy element. In this case we should also conclude that
 the NS is seen through a hole in the cloud R CrA,
unless the distance to the cloud was underestimated.
Thus, optical observations
of this object would enable one, for the first time, to probe 
chemical composition of surface layers of NSs
and to firmly evaluate their temperature. Obviously, the same
approach, i.~e., combined analysis of the X-ray and optical/UV
spectra with the use of the NS atmosphere models, can be applied to 
other similar objects.
\acknowledgements
We  thank Fred Walter, Werner Becker, Bernd Aschenbach,
Stuart Bowyer, and the referee for useful
discussions. The $ROSAT$ project is supported by the 
BMBF and the Max-Planck-Society. 
G.~P is grateful to the Max-Planck-Institut f\"ur Extraterrestrische 
Physik for hospitality; his work was partially supported by NASA grant 
NAG5-2807 and INTAS grant 94-3834. 
V.~Z. acknowledges Max-Planck fellowship and the INTAS grant.

\bigskip\noindent
\hrulefill

\bigskip\noindent
\large
This paper is accepted for publication in The Astrophysical Journal Letters, 1996
\end{document}